\newcommand\DIMPY{(C$_7$H$_{10}$N)$_2$CuBr$_4$}

\documentclass[prl,twocolumn,floatfix,preprintnumbers,amsmath,amssymb,superscriptaddress]{revtex4-1}

\usepackage{graphicx}
\usepackage{dcolumn}
\usepackage{bm}

\begin{document}

\title{Spectral and thermodynamic properties of a strong-leg quantum spin ladder}

 \author{D. Schmidiger}
 \affiliation{Neutron Scattering and Magnetism, Laboratory for Solid State Physics, ETH Zurich, Switzerland.}
 \author{P. Bouillot}
 \affiliation{DPMC-MaNEP, University of Geneva, CH-1211 Geneva,
Switzerland.}
 \author{S. M\"uhlbauer}
 \affiliation{Neutron Scattering and Magnetism, Laboratory for Solid State Physics, ETH Zurich, Switzerland.}
 \author{S. Gvasaliya}
 \affiliation{Neutron Scattering and Magnetism, Laboratory for Solid State Physics, ETH Zurich, Switzerland.}
  \author{C. Kollath}
 \affiliation{DPT-MaNEP, University of Geneva, CH-1211 Geneva,
Switzerland.}
 \author{T. Giamarchi}
 \affiliation{DPMC-MaNEP, University of Geneva, CH-1211 Geneva,
Switzerland.}
 \author{A. Zheludev}
 \email{zhelud@ethz.ch}
 \homepage{http://http://www.neutron.ethz.ch/}
 \affiliation{Neutron Scattering and Magnetism, Laboratory for Solid State Physics, ETH Zurich, Switzerland.}
\date{\today}

\begin{abstract}
The strong-leg $S=1/2$ Heisenberg spin ladder system \DIMPY\ is
investigated using Density Matrix Renormalization Group (DMRG)
calculations, inelastic neutron scattering, and bulk
magneto-thermodynamic measurements. Measurements showed qualitative
differences compared to the strong-rung case. A long-lived
two-triplon bound state is confirmed to persist across most of the
Brillouin zone in zero field. In applied fields, in the
Tomonaga-Luttinger spin liquid phase, elementary excitations are
attractive, rather than repulsive. In the presence of weak
inter-ladder interactions, the strong-leg system is considerably
more prone to 3-dimensional ordering.
\end{abstract}


\maketitle

In quantum magnets, the interplay between exchange and quantum fluctuations
leads to a host of novel phases, much richer than their classical counterparts.
In particular, correlations between the spins can be strongly
suppressed by quantum effects, leading to quantum \emph{spin liquid} phases
with properties quite different from those of any conventional ferro- or
antiferromagnet. Under magnetic fields these systems undergo quantum phase
transitions that are akin to Bose-Einstein condensation
\cite{Giamarchi2008}. Among the spin liquids,
antiferromagnetic (AF) Heisenberg $S=1/2$ ladders are the simplest,
yet perhaps the most important and extensively studied \cite{dagotto_ladder_review}.
They combine the essence of quantum magnetism with peculiar features that stem
from their one dimensional nature \cite{Giamarchibook}.
As a result, in applied fields they demonstrate a variety of scaling properties,
characteristic of the physics of one dimensional interacting quantum particles, the so called
Tomonaga Luttinger liquids (TLL). Understanding which key parameters of
the actual spin Hamiltonian control these universal features is a
formidable challenge that requires novel experimental and
theoretical approaches.

In recent years, a general theory of weakly coupled ladders under
strong magnetic fields has emerged \cite{Giamarchi1999}.
Considerable experimental progress in understanding
\emph{strong-rung} spin ladders was made through the study of the
compounds IPA-CuCl$_3$ \cite{Masuda2006,Garlea2007,Zheludev2007} and
BPCB
\cite{Lorenz2008,Ruegg2008,Klanjsek2008,Thielmann2009PRB,Thielmann2009,Savici2009}.
Particular attention was given to the field-induced quantum phase
transitions
\cite{Garlea2007,Zheludev2007,Lorenz2008,Klanjsek2008,Thielmann2009PRB}, and
the properties of the gapless TLL critical phase at intermediate fields
\cite{Klanjsek2008,Thielmann2009}.

In the case of the strong rung ladder, the spin gap in the absence
of a magnetic field is already present on each rung, protecting the
spin-liquid state from the leg exchange. A more subtle limit is provided by the strong leg (or weak
rung) ladder. In that case the existence of a spin liquid state is
far from obvious, and results \cite{dagotto_ladder_review} from an
Haldane gap mechanism \cite{haldane_gap}. This leads to some
similarities between the two limits but of course also to important
differences, in terms of the origin of the spin gap, excitation
spectrum, and the TLL mapping. On the experimental side, this
interesting problem remained elusive since only few studies are
available.

In this paper we report both experimental and theoretical studies of the prototypical strong-leg
spin ladder material DIMPY \cite{Hong2010,Schmidiger2011}. We
determine both its thermodynamic properties and the neutron
scattering spectrum, and show how to use these data to determine the
TLL parameters in the strong-leg case. A remarkable quantitative
agreement with DMRG calculations gives us a precise description of
the material, needed to understand long range ordering detected at
low temperatures.

\begin{figure}
\includegraphics[width=\columnwidth]{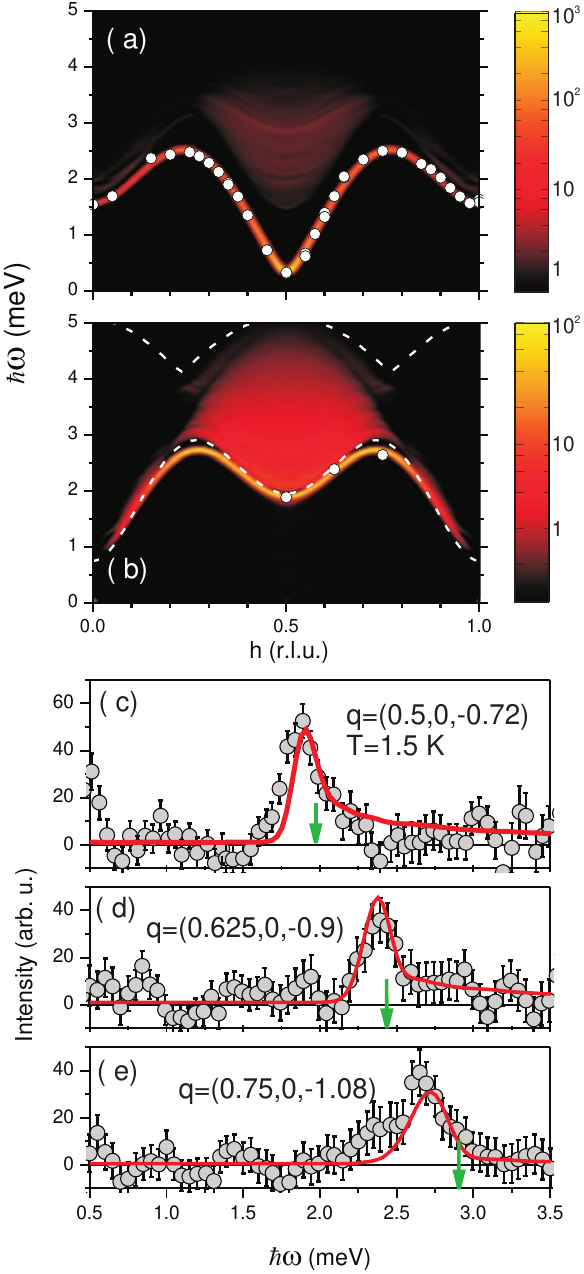}
\caption{(color online) Dynamic spin structure factor of DIMPY. (a) Antisymmetric
channel. The false-color plot shows the DMRG result\footnote{The oscillations are numerical artefacts.}.
Symbols are experimental data for single-triplon dispersion from
Ref.~\cite{Schmidiger2011}. (b) Symmetric channel. False-color plot
as above. Symbols are positions of peaks in inelastic neutron
scattering scans shown in (c)-(e) (symbols). The white dashed lines
in (b) are the limits of the two triplons continuum. The solid red
line in (c)-(e) is the DMRG result scaled by an arbitrary
factor and convolved with the resolution function of
the 3-axis spectrometer \cite{Popovici75}. The green
arrow in (c)-(e) is the lower edge of the two triplons
continuum.
\label{fig:spectrum}}
\end{figure}

The magnetic properties of DIMPY originate from ladders formed by
$S=1/2$ Cu$^{2+}$ ions, that run along the $a$ axis of the
monoclinic crystal structure \cite{Shapiro:2007}. We model this
compound by the AF Heisenberg two-leg spin ladder Hamiltonian
\begin{equation}
 \mathcal H = J_\mathrm{leg} \sum_{l,j}  {\bm S}_{l,j}
 \cdot {\bm S}_{l+1, j} + J_\mathrm{rung}\sum_l {\bm S}_{l,1} \cdot {\bm S}_{l,2}- g \mu_BH
 \sum_{l,j}S_{l,j}^z.\nonumber
 \label{eq:H}
\end{equation}
Here $J_\mathrm{leg}$ and $J_\mathrm{rung}$ are the couplings along
the leg and rung, respectively, $g\mu_\mathrm{B}H$ is the
uniform Zeeman field, and ${\bm S}_{l,j}$ are the spin operators
acting on site $l$ of the leg $j$ of the ladder. At $H=0$, the
ground state of DIMPY is a non-magnetic spin singlet separated from
the lowest-energy triplet excited states by an energy gap of
$\Delta=0.36$~meV \cite{Hong2010,Schmidiger2011}. Previous studies
suggested that the application of a magnetic field at $T\rightarrow
0$ leads to a quantum phase transition to the TLL state at
$H_{c1}=2.85$~T \cite{Hong2010}. Inelastic neutron scattering
measurements of the dispersion relation for triplon excitations
yielded an estimate of the ratio of exchange constants as
$J_\mathrm{leg}/J_\mathrm{rung}\sim 2.2$, through a comparison with
theoretical results obtained with the PCUT method \cite{Hong2010}. A
more detailed measurement over the whole Brouillon zone
confirmed that the spin Hamiltonian is symmetric
with respect to leg permutation \cite{Schmidiger2011}. This feature
allows one to conveniently describe the spin dynamics in terms of
separate \emph{antisymmetric} (leg-odd ``-'') and \emph{symmetric}
(leg-even ``+'') structure factors
\begin{equation}\label{equ:structfact}
\mathcal{S}^{(\pm)}(q,\omega)\propto\sum_\lambda|\langle\lambda|{\bm
S}_\pm(q)|0\rangle|^2\delta(\omega+E_0-E_\lambda),\nonumber
\end{equation}
respectively. Here $|0\rangle$ denotes the ground state of $\mathcal
H$ with energy $E_0$, ${\bm S}_\pm(q)=\sum_le^{-iqla}({\bm
S}_{l,1}\pm {\bm S}_{l,2})$, $a$ the lattice constant, and $\sum_\lambda$ is the sum over all
eigenstates $|\lambda\rangle$ of $\mathcal H$ with energy
$E_\lambda$. The two channels can be independently probed by
inelastic neutron scattering experiments.
\begin{figure}
\includegraphics[width=\columnwidth]{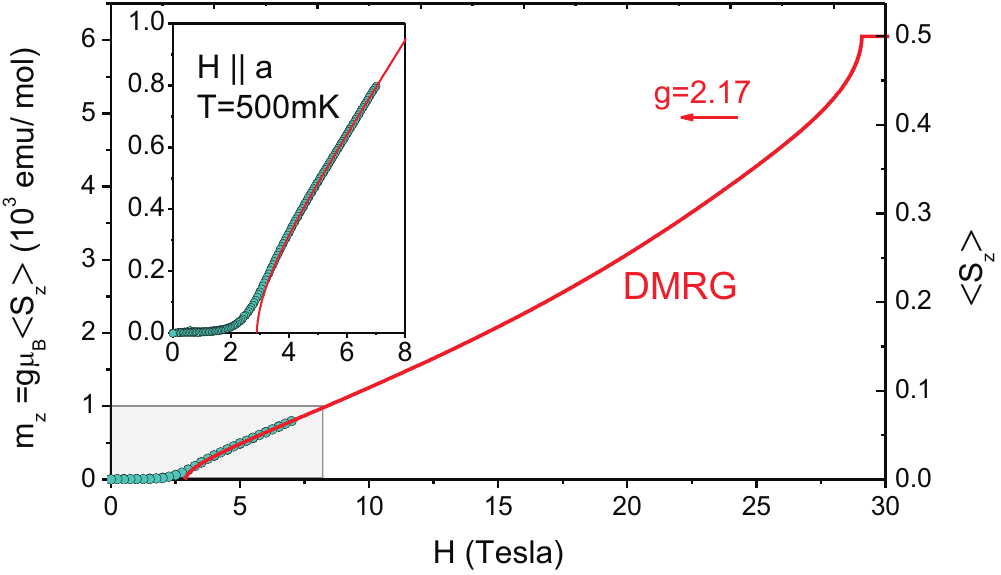}
\caption{(color online) Magnetization induced in DIMPY as a function
of applied field. Symbols are the experimental data obtained for $H\|a$
at $T=500$~mK. The solid line is a $T=0$ DMRG calculation. The inset
shows a blowup of the low-field region. \label{fig:magn}}
\end{figure}

To validate the spin Hamiltonian and to obtain a more accurate
estimate of the $J_\mathrm{leg}/J_\mathrm{rung}$ ratio, we fit the
experimental results of Ref.~\cite{Schmidiger2011} for the
\emph{full} single-triplon dispersion present in
$\mathcal{S}^{(-)}(q,\omega)$ with quasi-exact numerical results
adjusting both $J_\mathrm{leg}$ and $J_\mathrm{rung}$. The
calculations were performed using the time-dependent DMRG method
\cite{DaleySchollwoeck2004,White} (for specific details see
\cite{Bouillot2011}). An almost perfect agreement with experiment is
obtained over the whole Brillouin zone (BZ)
 with $J_\mathrm{leg}=1.42(6)$~meV and
$J_\mathrm{rung}=0.82(2)$~meV shown in Fig.~\ref{fig:spectrum}(a)
\footnote{See also the supplementary material.}. The excellent
agreement with data validates that $\mathcal H$ is a faithful
description of the system, and that additional terms (anisotropies,
Dzialoshinski-Moryia etc.) if present are extremely small. We obtain
$J_\mathrm{leg}/J_\mathrm{rung}=1.72(6)$ for DIMPY, which is notably
less than the value quoted in Ref.~\cite{Hong2010}. The main
difference occurs in $J_\mathrm{leg}$, and we attribute this
difference to the approximation within the PCUT
method~\cite{Hong2010} and to fitting in the whole BZ in our case.

The obtained exchange constants were used to calculate the symmetric
structure factor $\mathcal{S}^{(+)}(q,\omega)$. In the strong-rung
limit these excitations are attributed to multi-particle states with
an even number of triplons \cite{Barnes1993,Notbohm2007}. Assuming
no interactions between excitations, one expects to see a diffuse
continuum of two-triplon scattering with a \emph{maximum} of the
lower boundary at the center of the BZ $ka=\pi$. Interactions lead
to two-triplon bound states \cite{Sushkov1998}. In the strong rung
limit, these only exist below the continuum in a narrow range close
to the BZ-center. The actual calculated symmetric spectrum for DIMPY
is shown in Fig.~\ref{fig:spectrum}(b) and deviates from this
simplistic picture. Similarly to the isotropic point
($J_\mathrm{leg}=J_\mathrm{rung}$) \cite{knetter_ladder}, the
continuum has a local \emph{minimum} in the center of the BZ, where
most of the spectral weight is concentrated. A characteristic 'hat'
on top of the continuum can be identified. However, the most
prominent feature is a long-lived excitation below the boundary of
the continuum, stable across most of the BZ, at $0.8\cdot
2\pi\gtrsim ka\gtrsim0.2\cdot 2\pi$. Numerically, integrating the
singular and non-singular parts of the dynamic structure factor up
to 5~meV, we estimate that 56\% of the spectral weight is contained
in single-triplon excitations and 14\% in two-triplon bound states.

The theoretical results were tested in inelastic neutron experiments
at the TASP 3-axis spectrometer at Paul Scherrer Institute, using
the same deuterated single crystal samples and experimental
conditions as in Ref.~\cite{Schmidiger2011}. Typical constant-$q$
scans measured at $T=1.5$~K at several wave vectors that correspond
to $\mathcal{S}^{(+)}(q,\omega)$ are shown in
Fig.~\ref{fig:spectrum}(c-e) in symbols \footnote{To subtract the
background, we repeated the same scans at $T=50$~K and $T=100$~K,
where the magnetic contribution is expected to be wiped out. The
signal at high temperature was decomposed into $T$-dependent and
$T$-independent parts, assuming that the latter is due to phonons
and therefore scales with the Bose factor. From this analysis the
combined background was calculated for $T=1.5$~K and subtracted from
the data shown. All operations were performed point-by-point}. The
only adjustable parameter is an overall scale factor. The
quantitative agreement between theory and experiment is a
spectacular validation of our approach. In particular, it is
possible to experimentally separate the bound state from the
continuum. This is more delicate in strongly dimerized
compounds~\cite{tennant_two_magnon}, or those with large energy
scales \cite{Notbohm2007}.
\begin{figure}
\includegraphics[width=\columnwidth]{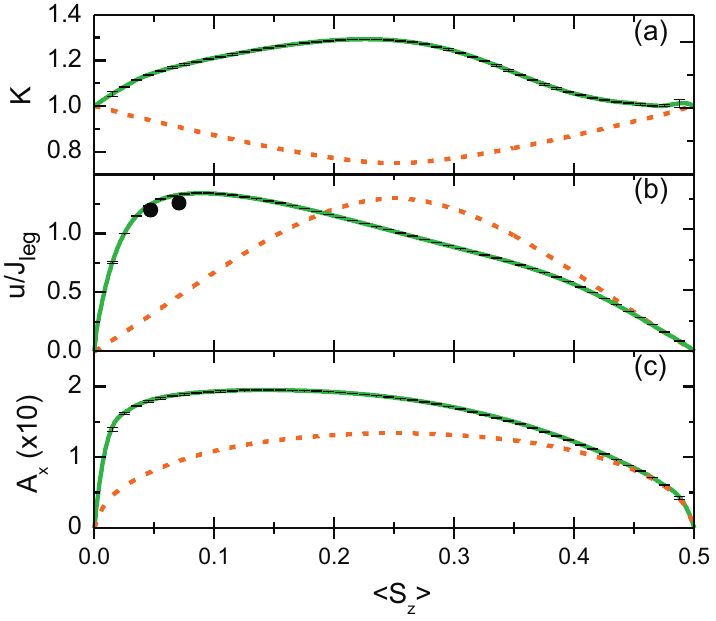}
\caption{(color online) TLL parameters as a function of field-induced magnetization.
DMRG results determined following Ref.~\cite{furusaki2000,Bouillot2011} for DIMPY with $J_\mathrm{leg}/J_\mathrm{rung}=1.72(6)$ are shown as solid lines and
the strong-rung coupling limit as dashed lines \cite{Bouillot2011}.
The symbols in (b) are extracted from the experimental heat capacity measurement. \label{fig:TLL}}
\end{figure}

The fitted Hamiltonian also allows us to interpret bulk magnetometric
experiments. The measured magnetization curve \footnote{The data
were collected on a Quantum Design Magnetic Properties Measurements
System MPMS-XL with an iQuantum $^3$He refrigerator. $g=2.17$ is
known from the independently measured gap energy and critical
field.}, for a field applied along the $a$ axis at $T=500$~mK is in excellent agreement with DMRG results as shown in Fig.~\ref{fig:magn}.
The small discrepancy in the very vicinity of $H_{c_1}$ being due to
finite-$T$ effects. The onset of magnetization signals the  gapless
TLL regime. Here, the low-frequency long wavelength correlation
functions and other properties are expected to have a universal form
determined by the so-called Luttinger parameter $K$, which defines
the powers of the algebraic correlations, and the velocity $u$ of
the linear excitation spectrum.

\begin{figure}
\includegraphics[width=\columnwidth]{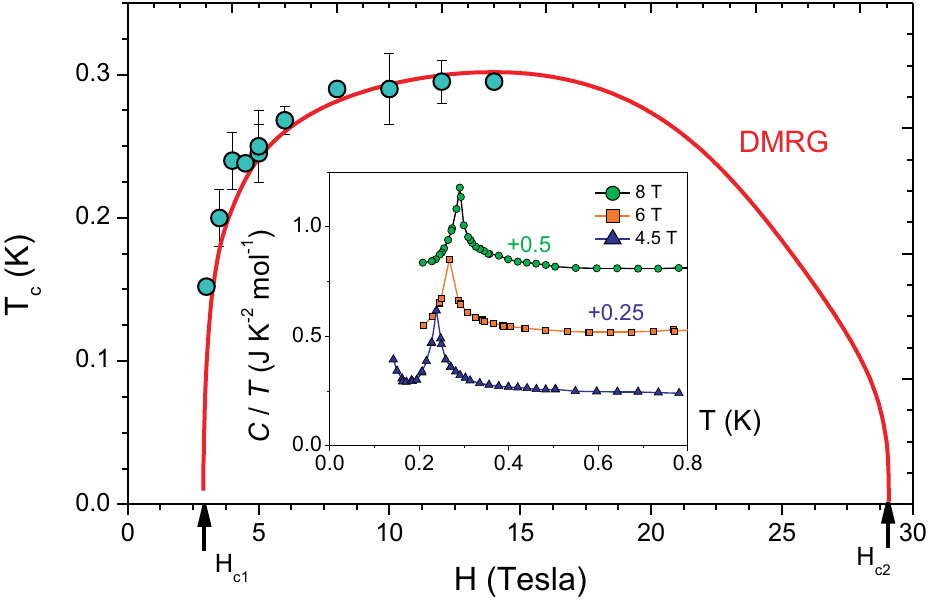}
\caption{ (color online) Inset: Magnetic specific heat measured in
DIMPY in fields applied along the $b$ axis.  Main panel:
Field-temperature phase diagram of DIMPY. The area between $H_{c1}$
and $H_{c2}$ is the ordered state. Circles are positions of lambda-
anomalies in specific heat. The solid line is the DMRG result using
the adjustable parameter $nJ'_\mathrm{MF}$.\label{phase}}
\end{figure}

In DIMPY, these field dependencies are markedly different from those
in the strong-coupling limit as shown in Figs.~\ref{fig:TLL}(a,b).
In the latter case, $K$ \emph{decreases} beyond $H_{c1}$, and
returns to unity at saturation $H_{c2}$. Throughout the TLL phase
$K<1$, and the elementary spin excitations (spinons) are
\emph{repulsive}. Not so in the strong-leg ladder. For DIMPY we see
that $K$ \emph{increases} beyond $H_{c1}$ and remains greater than
unity at higher fields. This signifies an \emph{attractive}
interaction between spinons \cite{Giamarchi1999}. In the direct
proximity of saturation at $H_{c2}$, $K\sim 1$, which corresponds to
non-interacting spinons. The velocity $u$ in DIMPY also behaves
quite differently compared to the strong-rung coupling case, showing
a strongly asymmetric behavior. This behavior of the velocity will
have a strong influence on numerous quantities defined by low-energy
excitations, such as the low energy continuum in the gapless phase.
As a consistency check, we estimated the velocity additionally from
the specific heat measurements discussed below, using the relation
$C(T)=\frac{\pi k_BT}{6u}$, where $C$ is normalized per spin
\footnote{Note the factor 2 difference compared to Eq.~(2) in
Ref.~\cite{Hong2010}. In our notation, specific heat is calculated
{\it per spin}, as is customary \cite{Giamarchibook},  and not per
rung. Our measured $C/T$ values are in good agreement with the
experimental data in  \cite{Hong2010}.}. This estimate (symbols in
Fig.~\ref{fig:TLL}b) are in good agreement with our calculated
velocity, in particular, considering that the determination by the
specific heat can be inaccurate, as detailed in \cite{Bouillot2011}.

TLL physics is endemic to one dimension. Ironically, one of the most
accurate ways to probe its properties is to study the {\it quasi}-1D
case of weakly coupled ladders. Inter-ladder interactions result in
three-dimensional long range AF ordering at a finite temperature.
Assuming unfrustrated and weak couplings, the problem can be treated
in the framework of the chain-mean field (MF) theory
\cite{Scalapino1975}. The characteristics of the ordered state are
entirely defined by the TLL properties of isolated ladders, with
only one added parameter: the effective inter-ladder coupling
constant $nJ'_{\mathrm{MF}}$. (The form suggests equal coupling
strength $J'_{\mathrm{MF}}$ to $n$ ladders). In particular, the
field dependencies of the ordering temperature $T_c$ is given by
Eq.~(2) of Ref.~\cite{Klanjsek2008}. In this formula the quantity
$A_x$ is the amplitude of AF correlations in isolated chains
(Fig.~\ref{fig:TLL}(c)). Combined with the field dependence of
$\langle S_z\rangle$, this gives us the field-temperature phase
boundary shown in a solid line Fig.~\ref{phase}.

DIMPY was previously hailed as an almost perfect 1D system that even
at $H>H_{c1}$ remains disordered \cite{Hong2010}.  In fact, more
careful specific heat measurements reveal a weak but well defined
lambda anomaly that appears for $H>H_{c1}$. This can be interpreted
as the onset of 3D long-range order. Typical spin specific heat data
collected in protonated samples for $H\|a$, are shown in
Fig.~\ref{phase} \footnote{The data were collected on a Quantum
Design Physical Properties Measurement System with a $^3$He-$^4$He
dilution refrigerator. The calculated Shottky contribution of
nuclear spin was subtracted. In the studied temperature range
lattice contribution to specific heat is totally negligible.}. At
each field, the putative ordering temperature $T_c$ was identified
with the peak position. It is plotted against field in symbols in
Fig.~\ref{phase} (right axis). The experimentally measured phase
boundary is in excellent agreement with the chain-MF prediction
assuming an un-frustrated inter-ladder coupling of
$nJ'_{\mathrm{MF}}=6.3~\mu$eV. This agreement lends credence that
the singularity seen in specific heat is indeed associated with the
3D ordering.

As MF neglects the quantum fluctuations between the ladders,
$J'_{\mathrm{MF}}$ may underestimate the real coupling $J'$. This
said, given almost the same inter-ladder MF coupling as in the
strong-rung material BPCB ($nJ'_{\mathrm{MF}}=6.9~\mu$eV
\cite{Klanjsek2008}), the ordering temperature is considerably
enhanced in the strong-leg case of DIMPY. This effect is principally
due to the rapid growth of transverse correlations as defined by
$A_x$, and their slow falloff due to the large $K$, showing again
the differences between the strong leg and strong rung limits.

This work is partially supported by the Swiss National Fund under
MaNEP and Division II. We thank  T. Yankova for her involvement in
the synthesis of DIMPY samples.

Note added: During the final stage of this work we became aware of
the study by Ninios {\it et al.}, arXiv:1110.5653v1, which contains
experimental data similar  to those shown in Fig.~\ref{phase}.


%

\end{document}